\documentclass{article}
\usepackage{spconf,amsmath,amssymb, graphicx,cite,booktabs, stix, printlen, tabularx, array, hyperref, url}
\usepackage{fancyhdr}

\fancypagestyle{firstpage}{
  \fancyhf{} 
  \fancyfoot[C]{\small{© 2024 IEEE. Personal use of this material is permitted. Permission from IEEE must be obtained for all other uses, in any current or future
media, including reprinting/republishing this material for advertising or promotional purposes, creating new collective works, for resale or
redistribution to servers or lists, or reuse of any copyrighted component of this work in other works.}} 
}

\title{Performance and Energy Balance: A Comprehensive Study of State-of-the-Art Sound Event Detection Systems}

\name{Francesca Ronchini$^{1}$,
       Romain Serizel$^{2}$
       }

\address{$^1$ Dipartimento di Elettronica, Informazione e Bioingegneria - 
              Politecnico di Milano, Milan, Italy \\
         $^2$Université de Lorraine, CNRS, Inria, Loria, Nancy, France \\
}

\begin{document}
\ninept

\maketitle

\thispagestyle{firstpage} 

\begin{abstract}
In recent years, deep learning systems have shown a concerning trend toward increased complexity and higher energy consumption. As researchers in this domain and organizers of one of the Detection and Classification of Acoustic Scenes and Events challenges task, we recognize the importance of addressing the environmental impact of data-driven SED systems. In this paper, we propose an analysis focused on SED systems based on the challenge submissions. This includes a comparison across the past two years and a detailed analysis of this year's SED systems. Through this research, we aim to explore how the SED systems are evolving every year in relation to their energy efficiency implications
\footnote{The data related to the submissions used for the analysis are available at \url{https://github.com/RonFrancesca/SED-carbon-footprint}}.

\end{abstract}

\section{Introduction}
\label{sec:intro}
Deep learning has yielded incredible achievements in a variety of audio processing applications, including Speech Recognition \cite{radford2023robust}, Machine Listening \cite{wang2023four, kim2023semi}, and Music Generation \cite{agostinelli2023musiclm, caillon2021rave}.
Despite the outcome of the remarkable results, the computational overhead of deep learning remains significant and continues to increase \cite{sevilla2022compute}. 
One particular concern is the substantial energy usage and resulting carbon footprint tied to the computational needs of deep learning. This situation is rapidly becoming unsustainable from both technical and environmental perspectives \cite{thompson2020computational, gupta2021chasing, schwartz2020green}. 
Until very recently, the environmental impact of deep learning models has largely been dominated by the persistent demand for high accuracy and effectiveness \cite{douwes2023quality}. In the last years, there has been a rise of concern about the environmental impact and energy consumption of deep learning within audio signal processing communities \cite{bannour2021evaluating, douwes2023quality,serizel2023performance, cerutti2019neural, parcollet2021energy}. 

However, comparing accurately the energy of different models, possibly trained on different sites is not straightforward~\cite{serizel2023performance}. There is no real consensus on the metric to be used and the relation between the different potential metrics (complexity, MACS, energy consumption) is often unclear. 
In the last years, different open-source Python packages have been introduced by the community in order to face this issue \cite{anthony2020carbontracker, rasley2020deepspeed, schmidt2021codecarbon, zhu2019thop}. Anyway, none of them provide a complete overview of the environmental impact. In fact, several factors must be taken into account when quantifying the energy consumption of deep learning algorithms 
\cite{serizel2023performance, douwes2023quality}.
    
\newcolumntype{C}[1]{>{\centering\arraybackslash}m{#1}}

\begin{table*}[th!]
\centering
\footnotesize
 \begin{tabular*}{\textwidth}{@{\extracolsep{\fill}}l|C{1.2cm}C{1.2cm}C{1.2cm}|C{1.2cm}C{1.2cm}C{1.2cm}|C{1.2cm}C{1.2cm}C{1.2cm}}
   & 	\multicolumn{3}{c|}{{System complexity $\downarrow$}} & \multicolumn{3}{c|}{Energy train (kWh) $\downarrow$} & \multicolumn{3}{c}{Energy test (kWh) $\downarrow$} \\

     & 25\% & Median & 75\% & 25\% & Median & 75\% & 25\% & Median & 75\% \\
     \toprule
    2022 Entries & 2200000 & 6676303 & 18903660 & 1.815 & 3.699 & 17.291 & 0.010 & 0.026 & 0.046 \\ 
    2023 Entries & 4804956 & 14662273 & 97176570 & 1.615 & 4.295 & 13.975 & 0.019 & 0.035 & 0.283 \\
 \bottomrule
 \end{tabular*}
 \caption{General comparison between DCASE 2022 and DCASE 2023 submissions entries. The table presents the median, 25th percentile, and 75th percentile values for system complexity, training energy, and test energy.}
 \label{tab:general}
\end{table*}

Starting from 2018, the objective of DCASE task 4 has been to investigate SED using a heterogeneous dataset featuring audio soundscapes with varying levels of label detail\cite{turpault2020training}. Throughout the years, alongside the ranking procedure, the evaluation of diverse submissions in practical scenarios has been crucial to gaining insights into the systems' performance  \cite{ronchini2022benchmark, turpault2019sound, serizel2019sound, ebbers2023post}. In DCASE task 4,  as in many other audio tasks, there has been a consistent trend of models steadily increasing in parameter complexity, frequently incorporating ensemble techniques. 
As also reported in Serizel et. al~\cite{serizel2023performance}, as organizers of the DCASE task 4, we recognized a responsibility to raise awareness about the carbon emissions and environmental implications associated with data-driven SED systems. 

In 2022, we asked participants to report the energy consumption of their systems both at training and test time \cite{ronchini2022benchmark}. We also introduced a new energy consumption metric \cite{ronchini2022description}, based on CodeCarbon toolkit \cite{schmidt2021codecarbon}. The metric has been introduced as an optional metric due to possible biases in terms of hardware used and fairness of comparison between systems. Since 2023, energy consumption reporting is mandatory. We also asked every participant to report energy consumption for 10 epoch of training of the baseline on their setup. This aims at normalizing the energy consumption metric by accounting for possible hardware disparities \cite{serizel2023performance}. We further asked participants to report an additional hardware-agnostic metric involving the computation of Multiply-Accumulate operations (MACs) for ten seconds of audio prediction.  We employed \textit{THOP: PyTorch-OpCounter} as a framework for MACs computation \cite{zhu2019thop}. 

This paper presents an analysis of the general evolution trend between 2022 and 2023. Following this initial overview, we focus on SED systems submitted in 2023. The metrics gathered in 2023 include hardware normalization and allow for a fairer comparison.
The goal of the analysis is to provide insights for achieving a better balance between performance and energy efficiency in SED system development.

\section{Analysis setup and evaluation metrics}
\label{sec:sec1}
The analysis is conducted on DCASE task 4 submissions in 2022 and 2023. For more information on system submissions, the reader is invited to visit the DCASE Challenge website \footnote{\label{foo:website}https://dcase.community/challenge2023/task-sound-event-detection-with-weak-labels-and-synthetic-soundscapes}.  
Within DCASE task 4, systems performance is evaluated with the polyphonic sound event detection scores (PSDS) \cite{bilen2020framework}. PSDS allows users to define parameter sets, defining customized scenarios under which SED systems are evaluated. For DCASE task 4, two scenarios are considered, as described in Ronchini et. al \cite{ronchini2022benchmark}. The PSDS is indicated throughout the paper as \textbf{PSDS\_1} when evaluated on scenario 1 and \textbf{PSDS\_2} when evaluated on scenario 2, regardless of the system.

From 2022, we propose a tentative, trivial energy weighted polyphonic sound detection score (EW-PSDS): 

\begin{equation}
    \mathrm{{EW-PSDS}} = \mathrm{PSDS} * \frac{\mathrm{kWh}_{\mathrm{baseline}}}{\mathrm{kWh}_{\mathrm{submission}}} 
    \label{eq:equa}
\end{equation}
where PSDS is the polyphonic sound event detection scores \cite{bilen2020framework}, $\mathrm{kWh}_{\mathrm{baseline}}$ is the energy consumption reported for the baseline, and $\mathrm{kWh}_{\mathrm{submission}}$ is the energy consumption of the submitted system \cite{ronchini2022description}.

The energy usage during both the training and test stages 
is reported using kWh (kilowatt-hour) as the unit.

This study presents an analysis of the relation between system performance metrics and energy consumption-related measures \footnote{\label{foo:add}Additional results are available at \url{https://github.com/RonFrancesca/SED-carbon-footprint}}. We report a subset of these metrics, along with MACs, system complexity (number of parameters of the deep learning model), and energy consumption. The energy usage is measured during both the training and inference stages.

\begin{figure}[t] 
    \centering
    \includegraphics[width=0.45\textwidth]{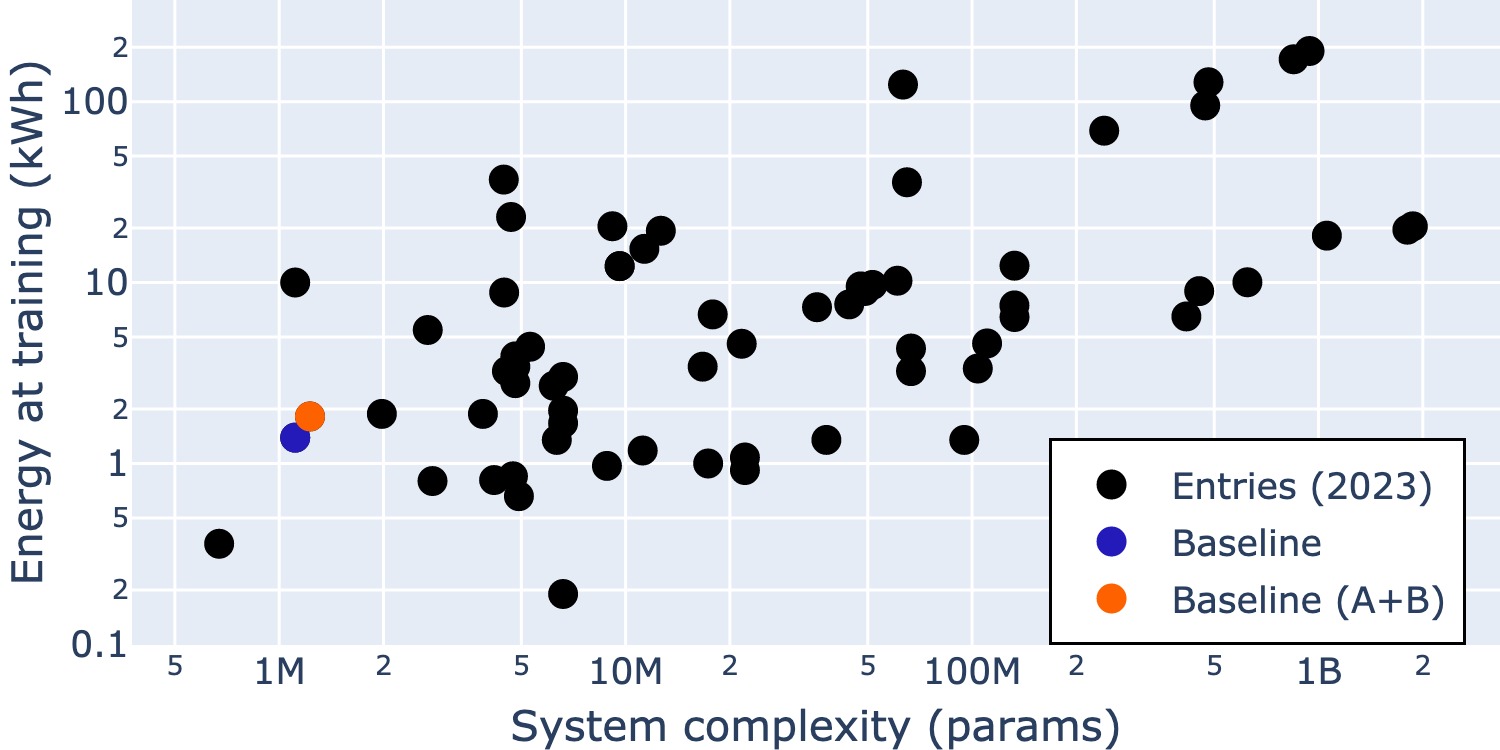} 
    \caption{Relation between system complexity and energy consumption at training for 2023 entries, compared with the two baselines systems.}
    \label{fig:sc_etr}
\end{figure}

\begin{figure}[t!] 
    \centering
    \includegraphics[width=0.45\textwidth]{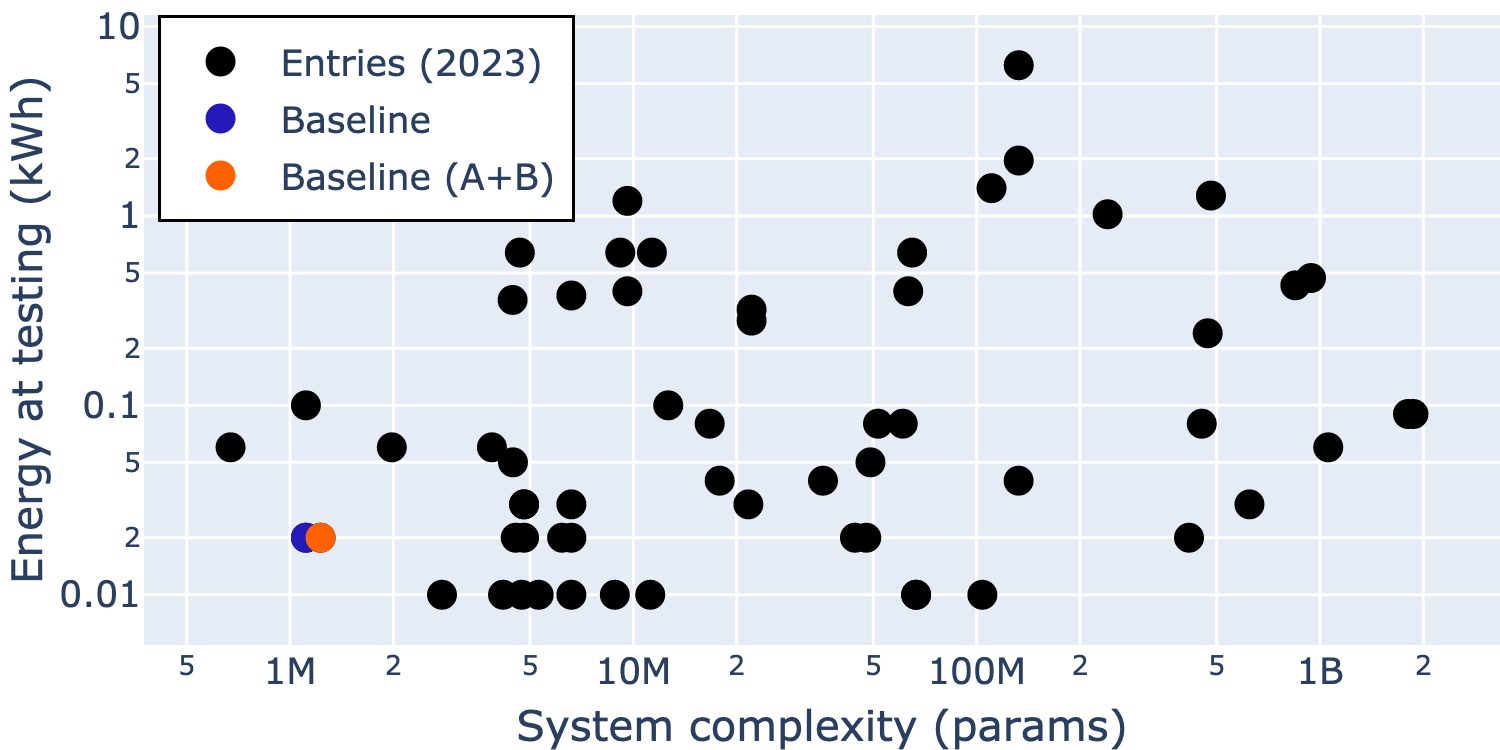} 
    \caption{Relation between system complexity and energy consumption at test for 2023 entries, compared with the two baselines systems.}
    \label{fig:sc_ete}
\end{figure}

\section{General comparison between DCASE 2022 and DCASE 2023 systems}
\label{sec:general}

This section analyzes all entries from 2022 and 2023 to understand how energy-related metrics have evolved over the past two years. In this study, we compare median, 25th percentile, and 75th percentile, for system complexity, energy consumption during training, and energy consumption during test. 
Quartiles and the median were chosen as measures of central tendency instead of the mean and standard deviation due to the presence of significant data variability. 
We conducted a filtering process in order to remove duplicate entries and entries with erroneous reported metrics. Initially, we had 101 total entries for 2022 and 123 total entries for 2023. After filtering, the analysis focused on 60 entries for 2022 and 64 entries for 2023.
Table \ref{tab:general} shows the results for system complexity, training energy consumption, and test energy for both 2022 and 2023 entries. Unsurprisingly, it can be observed a similar trend as the one discussed in Section \ref{sec:intro} for deep learning models. There is an inclination towards increased system complexity and energy consumption during both training and test.
While this is true, it is interesting to note that the energy consumption at training of the 75th percentile has remained stable or even decreased.

\begin{figure}[t!] 
    \centering
    \includegraphics[width=0.45\textwidth]{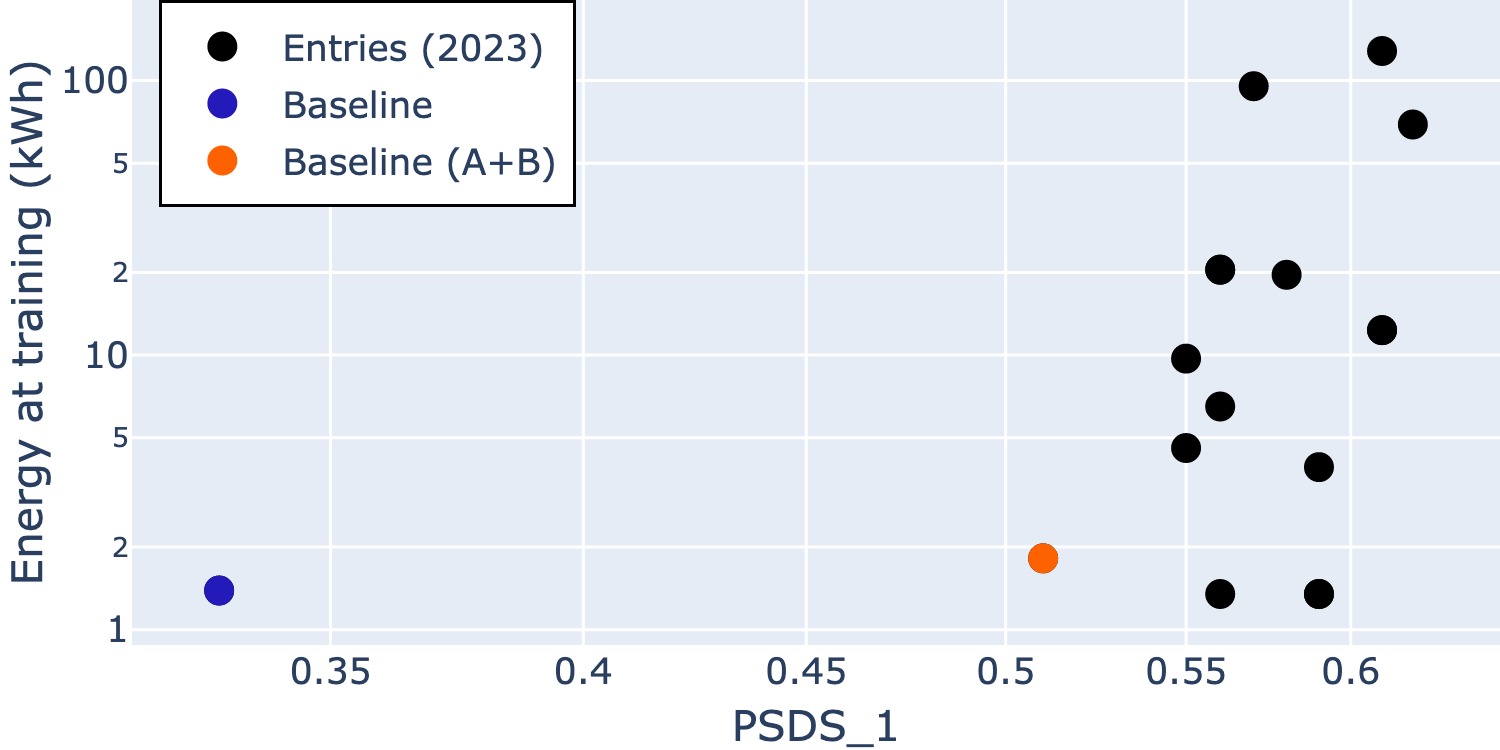} 
    \caption{PSDS\_1 and energy consumption at training for best 2023 systems, compared with the two baselines systems.}
    \label{fig:psds1_tr}
\end{figure}

\begin{figure}[t!] 
    \centering
    \includegraphics[width=0.45\textwidth]{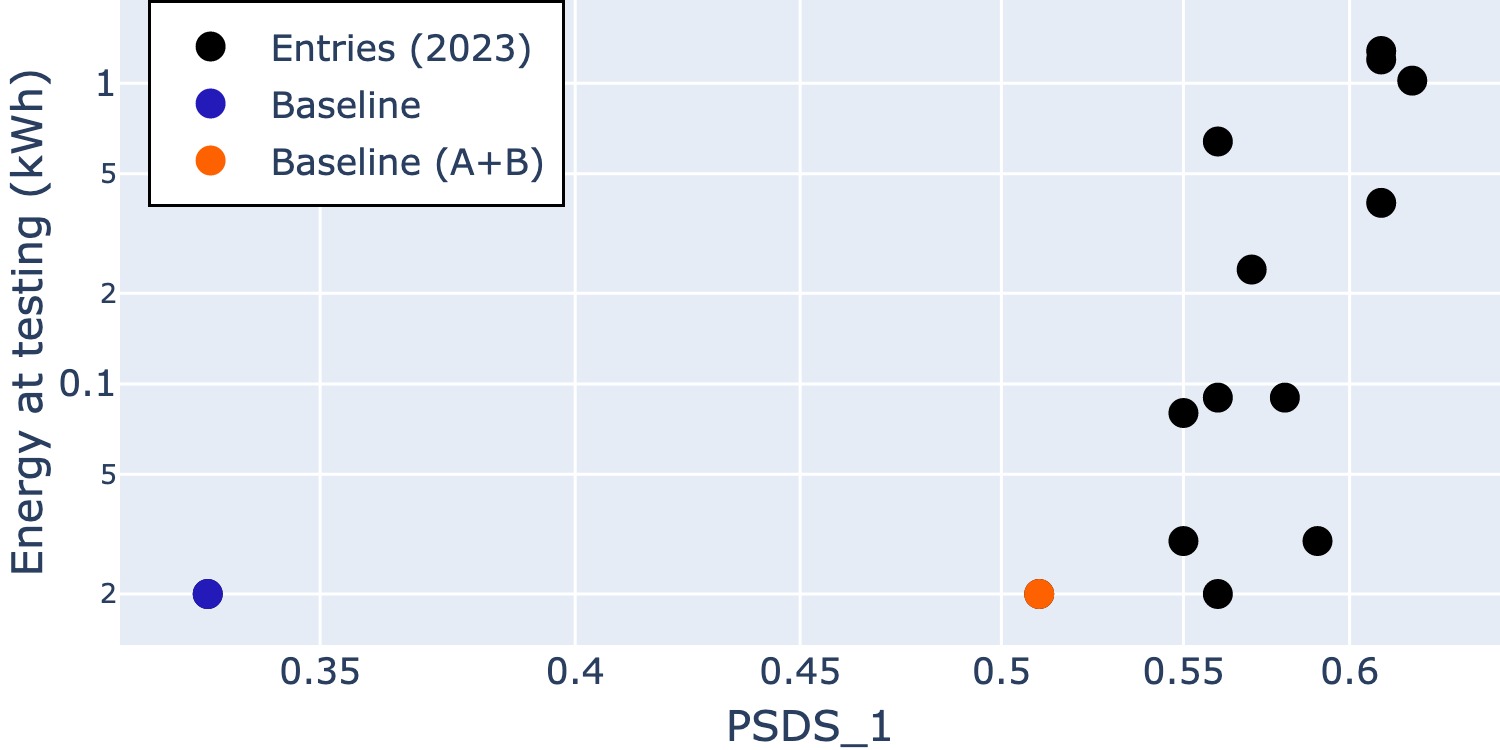} 
    \caption{PSDS\_1 and energy consumption at test for best performance 2023 systems, compared with the two baselines systems.}
    \label{fig:psds1_te}
\end{figure}

However, it's important to note that the energy consumption data for this general analysis is not normalized, which means we can hardly draw objective conclusions due to potential hardware biases. Additionally, not all 2022 submissions provided energy consumption values. This initial study is only presented to observe general trends. To ensure fairness and a more insightful study, the rest of the analysis is exclusively focused on results related to 2023 submissions.

\section{Relation between system complexity, MACs, and energy consumption}
\label{sec:macs}

\begin{figure}[t] 
    \centering
    \includegraphics[width=0.45\textwidth]{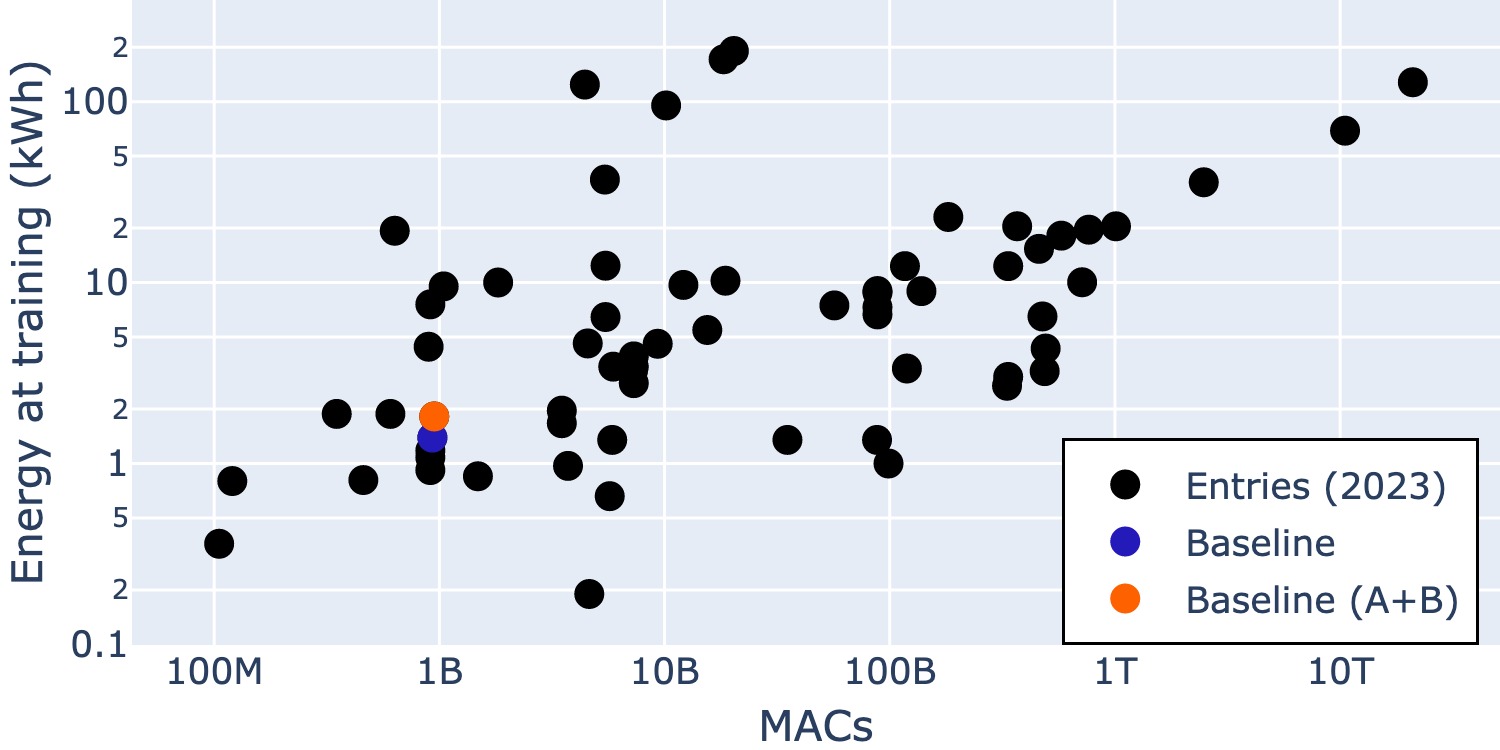} 
    \caption{Relation between MACs and energy consumption at training for 2023 entries, compared with the two baselines systems.}
    \label{fig:macs_tr}
\end{figure}

\begin{figure}[t] 
    \centering
    \includegraphics[width=0.45\textwidth]{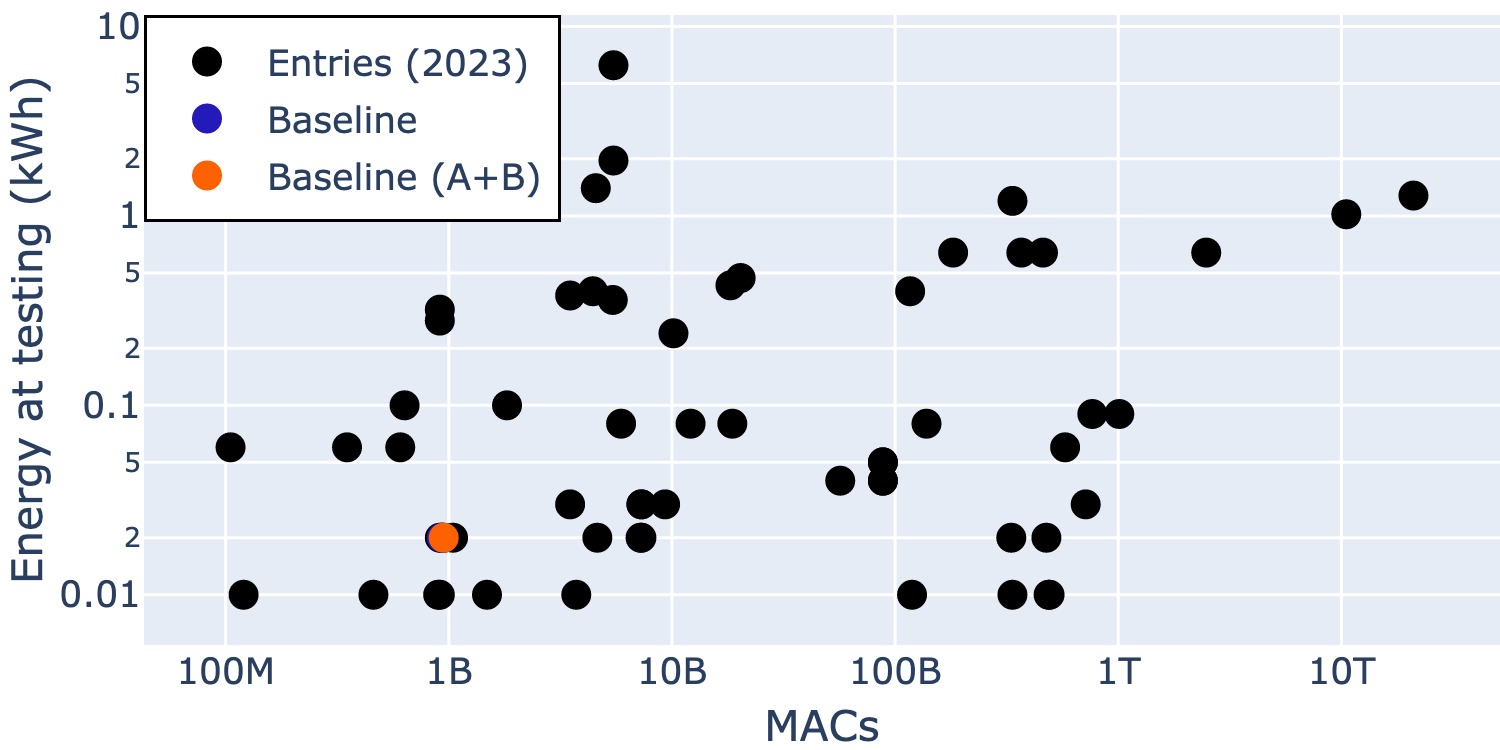} 
    \caption{Relation between MACs and energy consumption at test for 2023 entries, compared with the two baselines systems.}
    \label{fig:macs_te}
\end{figure}

In this section, we present an analysis of the MACs and system complexity in relation to energy consumption during the training and test phases of the DCASE 2023 submissions. To ensure a fair comparison among systems, energy consumption has been normalized by the baseline energy consumption \cite{serizel2023performance}. 

Figure \ref{fig:sc_etr} illustrates the correlation between system complexity and energy consumption during the training stage, Figure \ref{fig:sc_ete} presents the corresponding relation for energy consumption during the test stage. Figure \ref{fig:macs_tr} and Figure \ref{fig:macs_te} show the relation between MACs and energy consumed at the training phase, and the energy consumed at the test phase, respectively. On all the plots, we also report
the performance of the two baselines proposed for the challenge. \textit{Baseline} indicates the simple baseline which does not use external dataset or embeddings, while \textit{Baseline (A+B)} stands for the baseline using Audioset as external dataset \cite{hershey2021benefit} and BEATs embeddings \cite{chen2022beats} \footnote{See also the task webpage \textsuperscript{\ref{foo:website}} for more details.}.

From the results, it is possible to observe that MACs correlate slightly more with energy at training than system complexity. It is surprising that MACs correlate better with training energy than test energy, while the other way around would be expected as MACs are computed at test time. This difference might be related to different system architectures but this would have to be verified in extensive experiments. 

The relation between system complexity with the energy consumed by the system at both training and test phases is not straightforward. These three different metrics independently are insufficient to provide a comprehensive understanding of the system's footprint. In fact, as an example, from Figure \ref{fig:sc_etr}, there are systems with a complexity of 5M parameters that consume more energy than systems with a complexity of 100M parameters. Similar observations apply to MACs counts and energy consumption. However, for simplicity sake, we will focus the analysis mainly on energy consumption for the rest of the paper.


\begin{table*}[t!]
\centering
 \begin{tabular}{l|cc|cc|cc||cc|cc|cc}
 
       & \multicolumn{2}{c|}{{System complexity}} & \multicolumn{2}{c|}{MACs  } & \multicolumn{2}{c||}{Energy train (kWh)}   & \multicolumn{2}{c|}{{System complexity}} & \multicolumn{2}{c|}{MACs  } & \multicolumn{2}{c}{Energy train (kWh)}  \\
 
     & Max & PSDS\_1 & Max & PSDS\_1 & Max & PSDS\_1 & Max & PSDS\_1 & Max & PSDS\_1 & Max & PSDS\_1 \\
    \toprule
    All & 1B & 0.59 & 492 B & 0.59 & 23.00 & 0.59 & 1B & 0.62 & 21 T & 0.62 & 190.00 & 0.62\\
    25th & 5 M & 0.55 & 912 M & 0.55 & 0.99 & 0.55 & 25 M & 0.61 & 8 B & 0.58 & 4.59 & 0.60 \\
    Median & 6 M & 0.59 & 4 B & 0.55 & 2.33 & 0.56 &67 M & 0.61 & 72 B & 0.60 & 9.34 & 0.60 \\
 \bottomrule
 \end{tabular}
 \caption{The table presents PSDS\_1 when system complexity MACS and training energy are thresholded to the median value or the 25th percentile. The left side is related to no-ensemble systems, the right is related to ensemble systems.}
 \label{tab:thresholding}
\end{table*}

\begin{figure}[t!] 
\centering
    \includegraphics[width=0.45\textwidth]{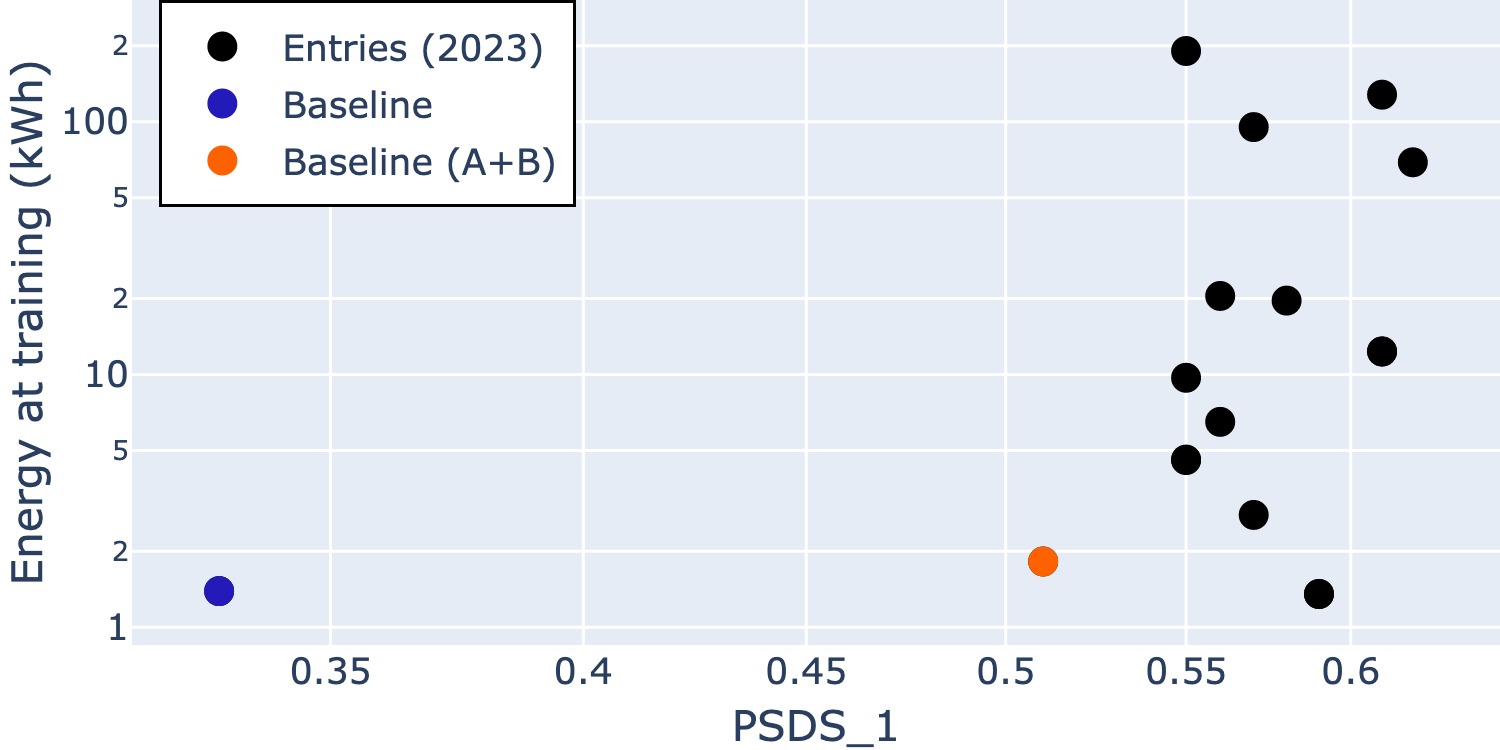} 
    \caption{Relation between PSDS\_1 and energy consumption at training for the best ensemble systems.}
    \label{fig:ens_energy}
\end{figure}

\begin{figure}[t] 
    \centering
    \includegraphics[width=0.45\textwidth]{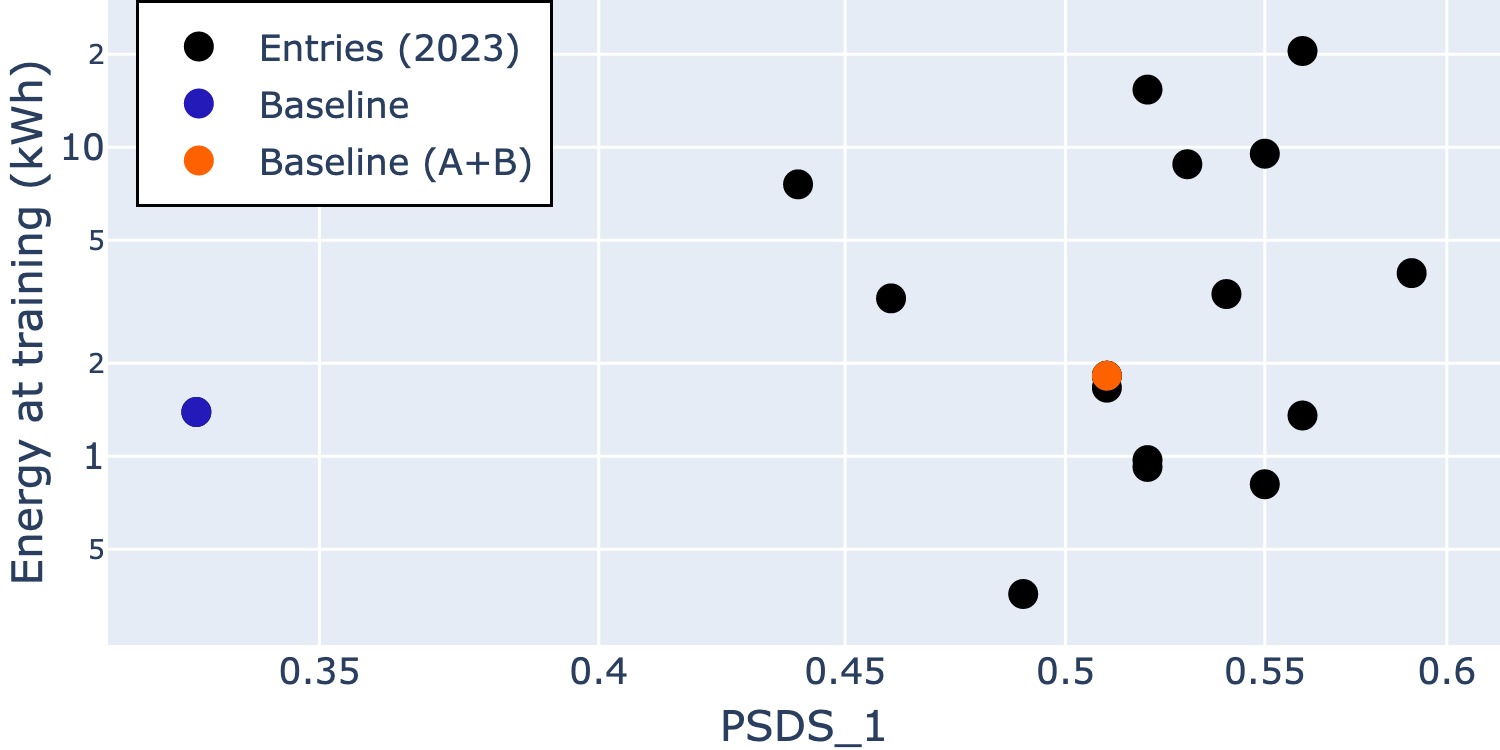} 
    \caption{Relation between PSDS\_1 and energy consumption at training for the best not-ensemble systems.}
    \label{fig:no_ens_energy}
\end{figure}

\section{Relation between performance and energy consumption}
\label{sec:psds}

This section presents findings regarding the relation between performance and energy consumption. We focus on the top 15 systems in terms of PSDS\_1 performance along with training and test energy consumption. 
If the same team has more than one submission with values within a range of 1-2 points in terms of PSDS\_1, we selected only one submission per team, the best in terms of PSDS\_1. We removed duplicate entries for each team based on MACs, keeping the best-performing system for each team. \footnote{A similar analysis for PSDS\_2 is reported on the additional results.}

\begin{figure}[ht] 
    \centering
    \includegraphics[width=0.45\textwidth]{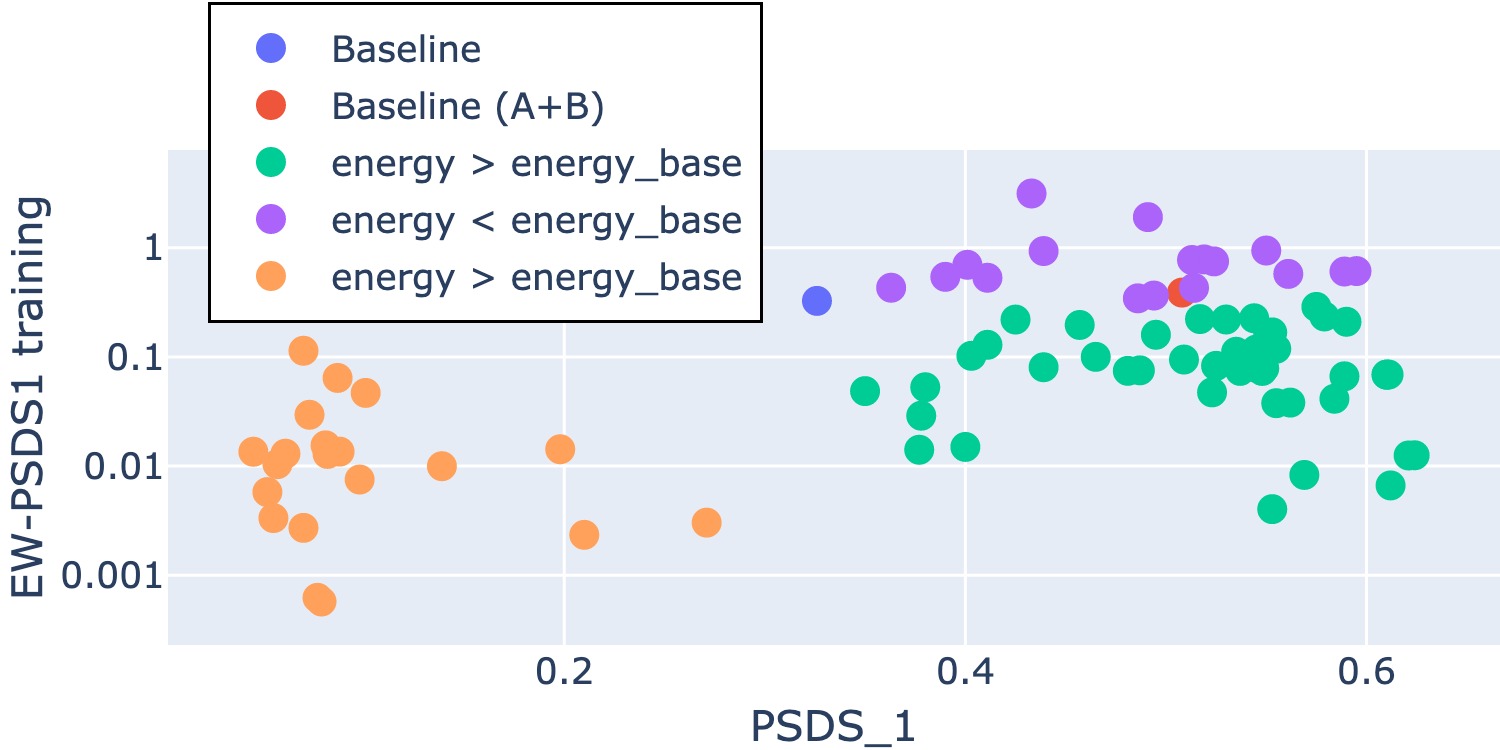} 
    \caption{Relation between PSDS\_1 and energy consumption weighted at training for 2023 entries, compared with the two baselines systems.}
    \label{fig:ew_psds1}
\end{figure}

Figure \ref{fig:psds1_tr} and Figure \ref{fig:psds1_te} illustrate the performance metrics PSDS\_1 in relation to training energy consumption, and test energy consumption, respectively. The figures highlight the diversity in terms of system efficiency. Particularly for PSDS\_1, some systems manage to outperform the baseline results while consuming less energy. Additionally, should be noted that the top-performing systems are not the systems that consume the most energy.  The same observation holds true for energy consumption during test. In fact, from the results, it appears that energy consumed for training and test have a similar distribution. In the remainder of the paper, we will focus on energy at training. We know that energy at test is the most important in terms of the footprint of deployed systems yet it also depends on the hardware target and is difficult to tackle in the current challenge setup. Additionally, according to previous observations energy at training can be considered as a first (gross) indicator of what would happen at the test (even though many other factors are involved). 

\section{Comparison between ensemble/non-ensemble systems}
\label{sec:ense}
This section analyzes the relation between PSDS\_1 and energy consumption for ensemble and not-ensemble systems. We still focus on the 15 best systems in terms of PSDS\_1. 

Figure \ref{fig:ens_energy} and Figure \ref{fig:no_ens_energy} show the relation between PSDS\_1 and energy consumption at training for the best ensemble-based submissions and for the top not-ensembled-based entries, respectively. What can emerge from this analysis is that an ensemble is useful at combining systems that alone are not so good in achieving decent performance (this is not so efficient in terms of energy but these not-so-good systems are already expensive) while a single system can provide a lighter alternative to reach good performance anyway. In fact, the best not-ensemble system is able to achieve a PSDS\_1 score of 0.58, while the best ensemble system scores 0.62 for PSDS\_1. 

\section{Relation between EW-PSDS and PSDS}
\label{sec:weight}

Figure \ref{fig:ew_psds1} shows the relation between PSDS\_1 and EW-PSDS. In particular, we can image the plot divided into four areas: bottom left corner, showing the system that were outperformed by the baseline and still had potentially higher consumption; bottom right corner: systems that outperform the baseline but consume more energy; top right: systems that outperform the baseline with limited energy consumption increase. The area to which we should aim is the top-right corner, which includes a system able to right high performance, not underestimating the environmental impact they are going to have. Unfortunately, as the figure shows, they are still minorities. 
This Figure highlights how most of the systems required a higher quantity of energy at training compared to the baseline. Some systems of the top right and the bottom right have pretty close PSDS\_1 while having pretty different energy consumption. 
The systems able to outperform the baseline necessitating less energy at training are three different submissions of the team \textit{Chen\_CHT} \cite{chen2023sound}.

\section{Thresholding based on energy consumption}
\label{sec:sec5}

The last part of the analysis evaluates how much the performances degrade if a footprint cap is set. In order to do so, we define a threshold related to the system complexity, MACs, and energy consumption. 
For each threshold, we select only the systems that have a lower value of the threshold. For example, when considering the median of the energy consumption at training, we selected the systems with a lower energy consumption than the median consumption and reported the best PSDS\_1 score. We applied a similar approach for MACs and system complexity. Due to space limitations, the same analysis for the 75th percentile is reported in the additional results, considering additional metrics \textsuperscript{\ref{foo:add}}. 
Table \ref{tab:thresholding} reports the degradation of the PSDS\_1 for the systems that have been thresholded according to the different metrics. The left side reports the results for non-ensemble systems, while the right reports the results for ensemble systems. The PSDS performance remains rather stable regardless of the threshold cap while the complexity, MACs and energy consumption are substantially decreased. This is even clearer where considering ensembling. This indicates that we are spending a large amount of energy and computation to increase the performance only marginally. 

\section{Conclusions}
\label{conclusions}

This paper presents a comprehensive examination of energy usage and its correlation with various metrics applied to systems participating in the DCASE task 4 during the years 2022 and 2023. The findings make it evident that relying on a single metric is insufficient for accurately measuring a system's footprint. The paper also highlights that systems consuming the most energy (or having the most MACs) do not necessarily outperform less computationally expressive systems. These observations highlight the pressing need for metric(s) capable of taking into account various factors to accurately estimate the energy consumption of deep learning models while taking into account the task-wise performance of the systems. This would be the first important step to effectively design sustainable SED systems.
\vfill\pagebreak


\newpage
\bibliographystyle{IEEEbib}
\bibliography{refs}

\end{document}